\documentclass[manuscript,screen]{acmart}
\AtBeginDocument{%
  \providecommand\BibTeX{{%
    \normalfont B\kern-0.5em{\scshape i\kern-0.25em b}\kern-0.8em\TeX}}}

\acmConference[Perhaps CHI]{Make sure to enter the correct
  conference title from your rights confirmation emai}{Somewhere}{Sometime}
\begin{document}

\title{Capital and CHI: Technological Capture and How It Structures CHI Research}


\author{Eric Gilbert}
\affiliation{
  \institution{University of Michigan}
  \streetaddress{\_}
  \city{\_}
  \country{USA}}
\email{eegg@umich.edu}

\renewcommand{\shortauthors}{ }

\begin{abstract}
This paper advances a theoretical argument about the role capital plays in structuring CHI research. We introduce the concept of \textit{technological capture} to theorize the mechanism by which this happens. Using this concept, we decompose the effect on CHI into four broad forms: technological capture creates \textit{market-creating, market-expanding, market-aligned,} and \textit{externality-reducing} CHI research. We place different \textsc{CHI} subcommunities into these forms---arguing that many of their values are inherited from capital underlying the field. Rather than a disciplinary- or conference-oriented conceptualization of the field, this work theorizes CHI as tightly-coupled with capital via technological capture. The paper concludes by discussing some implications for CHI.

\end{abstract}




\maketitle

\vspace{2pt}
\section{Introduction}
\vspace{2pt}

\begin{quote}
\emph{Growth, accumulation, and concentration of capital bring in their train an ever more detailed subdivision of labour, an ever greater improvement of old machines, and a constant application of new machine---a process which goes on uninterruptedly, with feverish haste, and upon an ever more gigantic scale.}
\\
--- Marx, \emph{Wage Labour and Capital}, 1847 \cite{marx1902wage}
\end{quote}

\vspace{10pt}

\begin{figure*}[!t]
    \centering
    \includegraphics[trim={2.9cm 3.7cm 3.1cm 3.7cm},clip,width=1\textwidth]{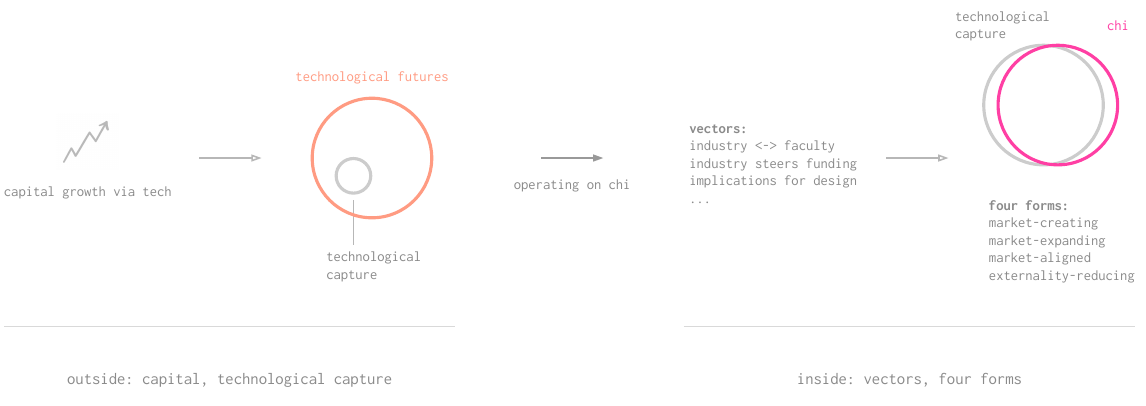}
    \caption{Structural overview of technological capture and CHI. Capital captures a set of technological futures aligned with its interests. CHI tightly couples to technological capture through a variety of vectors, resulting in four forms of CHI research.}
    \label{fig:capture-chi}
\end{figure*}

\noindent 
This paper advances a theoretical argument about the role capital plays in structuring CHI research.\footnote{We focus the argument on CHI the conference. Some of the arguments may extend to HCI more broadly.} We begin with a review of academic work on capital, foregrounding capital growth and interactions between modern tech and financial markets. Next, this paper contributes the concept of \textit{technological capture} to reason about how capital structures technology, and by extension, CHI research. The term is an analogy to regulatory capture \cite{dal2006regulatory}---when industries capture the agencies meant to regulate them. For example, in the United States, banks have successfully lobbied Congress to roll back protections put in place after 2008 to limit systemic collapse \cite{hill}. Similarly, technological capture is when capital captures technological futures, foreclosing on those unaligned with its interests. One argument this paper aims to make is that CHI has tightly-coupled with technological capture through a variety of different vectors---both present and historical (see Figure \ref{fig:capture-chi}).

We use frame analysis to decompose technological capture's effect on CHI into four broad forms: technological capture has produced \emph{market-creating}, \emph{market-expanding}, \textit{market-aligned}, and \emph{externality-reducing} types of CHI research. Market-creating research holds the potential to create new products and firms. Market-expanding research expands or retains customer bases. Market-aligned research puts technology into particularly lucrative markets. Finally, externality-reducing research helps to mitigate negative externalities caused by capital's interaction with technology. We place different \textsc{CHI} subcommunities into these forms: for example, technical HCI's strongly-held value of \textit{novelty} is, we argue, inherited from capital's interest in creating new products and companies (i.e., market-creating). Adopting an externality-reducing perspective: the research question ``Can a novel configuration of people and tech reduce misinformation online?'' may be better understood as ``Can a novel configuration of people and tech limit the likelihood that a government regulates tech platforms?''

This work follows in a line of reflexive theoretical work at CHI (e.g., \cite{alkhatib2021live,bardzell2013critical,blevis2018seeing,dourish2006implications,selbst2019fairness,zimmerman2007research}), as well as calls for greater engagement with market processes (e.g., \cite{lin2021techniques,lindtner2020prototype}). Rather than a disciplinary- or conference-oriented conceptualization of the field (e.g., \cite{grudin,big}), this work theorizes CHI as tightly-coupled with capital via technological capture. This work advances a theoretical position as a type of HCI contribution \cite{wobbrock2016research}; it is also in dialogue with applied work on capital's influence on research \cite{gansky2022counterfacctual,group2019patron,vertesi2023divesting,whittaker2021steep,young2022confronting}. 

The present work's primary purpose is to advance theoretical concepts, and is largely explanatory, rather than normative. However, after engaging with  counterarguments to the theoretical work, we conclude by reflecting on what a capital framing means for CHI research and for its corresponding technological innovation. Perhaps paradoxically, it may be the case that the most innovative, disruptive, and novel kinds of technologies and research are those that lie outside technological capture.



\section{Capital}
Next, we review capital and foreground concepts related to capital growth. We also connect historical work on capital to modern interactions between capital and tech through financial markets and VCs. Both help build Figure \ref{fig:capture-chi}'s left-side in the main argument of this paper. We conclude with a discussion of recent work exploring the impact of capital on computing research.

\vspace{8pt}
\noindent
\textit{\textbf{Capital.}} Capital is wealth, property, or assets the confer benefit to their owner. For example, a tech firm's capital might include: the code that runs its products; its owned infrastructure, including its data centers; its intellectual property (IP), in the form of patents, trademarks, and copyright, etc.; its financial reserves and investments in other companies, among many others. Capital is also held by individuals and families, and transmitted between generations as inherited wealth \cite{beckert2008inherited,piketty2015wealth}. Capital has been strongly implicated in the violence of colonialism (e.g., \cite{harris2004did}) and racism \cite{moten2003break}\cite[pg. 196]{piketty2014capital}\cite{sacerdote2005slavery}---at societal scales.

Capital and capitalism have have been studied from a variety of economic (e.g., \cite{solow1956contribution}) and sociological (e.g., \cite{granovetter2018sociology}) perspectives. In \textit{Capital in the Twenty-First Century} \cite{piketty2014capital}, French economist Thomas Piketty presents a historical account of capital over multiple centuries. Piketty's overall approach is a modern economic one. However, his historical review of the study of capital is helpful as framing: he contrasts Marx's body of work (the ``apocalyptic'' view in his language) with Kuznets's work a century later (the ``fairy tale''). 

Often better known for his political treatise \emph{The Communist Manifesto} \cite{marx1967communist}, Marx devoted a great deal of his life to theoretical, economic analyses of capitalism. Piketty writes:

\begin{quote}
\textit{His [Marx's] principal conclusion was what one might call the ``principle of infinite accumulation,'' that is, the inexorable tendency for capital to accumulate and become concentrated in ever fewer hands, with no natural limit to the process.} \cite[pg. 11]{piketty2014capital}
\end{quote}

\noindent
Marx believed that this ``infinite accumulation'' would at some point degrade capitalism's ability to produce growth, or lead to rebellion by workers who would ultimately own nothing. Piketty concludes, as many modern scholars have, that Marx's ``dark prophecy'' did not materialize because Marx failed to account for technological progress, among other things. (Piketty himself concludes, via his own empirical work, that wealth may continue to concentrate absent geo-political shocks or interventions.)

Piketty contrasts Marx with the work of Simon Kuznets \cite{kuznets2019economic}, written a century later during the Cold War. Kuznets, along with economist Robert Solow \cite{solow1956contribution}, argued that the benefits of economic growth would be spread across societal strata---and not, like Marx argued, inevitably concentrate in an ever smaller number of people. Kuznets explained his view with a famous aphorism: ``Growth is a rising tide that lifts all boats.'' Piketty notes that the Kuznets/Solow worldview held ``considerable influence ... in the 1980s and 1990s'' \cite[pg. 14]{piketty2014capital}; notably, this is the time during which the tech industry (and CHI itself) began in earnest.

\vspace{8pt}
\noindent
\textit{\textbf{Capital growth.}} A fundamental attribute of capital is accumulation and growth: capital continually seeks growth by producing \textit{rates of return} \cite{solow1963capital}. Capital seeks a rate of return (denote it $r$, as Piketty does), such that after time passes, capital currently worth total wealth $W$ will be worth $W \times (1+r)$. Capital produces rates of return in part by locating and exploiting \textit{surplus value} \cite{marx2004capital}. This is the difference between what a product is worth when you sell it, and what it costs to produce.

\vspace{8pt}
\noindent
\textit{\textbf{Tech and capital growth.}} Tech offers unique growth opportunities for capital. For example, Apple has produced an annualized total return of $r = $ 25.1\% over the last 15 years for its shareholders \cite{forbes}. An Apple shareholder with wealth $W$ invested in the company in 2008 would have $W \times 1.251$ in 2009, $W \times 1.251^2$ in 2010, and $W \times 1.251^{15} = W \times 28.76$ in 2023. They would have multiplied their wealth over 28-fold in 15 years.

Infrastructures based on code can infinitely copy and distribute products around the world, at diminishing marginal costs \cite{callon2007market,lessig2009code,terranova2000free}. These infrastructures minimize the negative term in the surplus value equation. When tech infrastructures alone are insufficient, capital seeks minimally expensive solutions. For example, OpenAI used Kenyan workers paid less than \$2/hour to train AI systems to be less ``toxic'' \cite{time}; Facebook outsources its difficult and traumatic content moderation decisions (a problem it has been unable to solve with algorithms alone) to low-wage workers in the U.S. and the Philippines \cite{verge}.

\subsection{Modern Tech, VCs, and Financial Markets}
In the United States, where most major tech firms are founded and headquartered, the ideology of ``shareholder primacy'' holds that tech firms \textit{exist} to supply returns to their owners and shareholders\footnote{There are a handful of exceptions---e.g., Anthropic was founded as a B-corp, or benefit corporation---however, they are a tiny fraction. Those firms also have unclear rules and impact.} \cite{smith1997shareholder,stout2001bad}. Explicitly, their reason for existence is to continually multiply capital investment for shareholders. They interact with complex streams of capital to do this.

Different tech firms take different trajectories, but at a high level, many tech firms interact with capital as follows: 1) initially, as privately-owned companies that seek investment from ``angel investors''\footnote{Angel investors are wealthy individuals that allocate capital to very early-stage companies.} and venture capital firms (VCs) \cite{conti2013show,denis2004entrepreneurial}; and 2) as perhaps, ultimately, public companies seeking capital investment in public markets (e.g., Meta, Alphabet, etc.)\footnote{This progression isn't always linear, as Twitter/X  demonstrated, going from a publicly-listed company back to a privately-held one.} \cite{pollman2019startup}. A startup might initially seek investment from friends, family, and angel investors. Next, it seeks investment in ``seed rounds'' and escalating ``Series A,'' ``Series B,'' etc. rounds---perhaps totaling hundreds of millions of dollars. Later, once the firm has become established, it may go public, and accept capital from investors all over the world. For example, Airbnb raised \$6.4B spread over 30 rounds \cite{airbnb}, starting with a \$20\textsc{K} investment from the influential startup incubator YCombinator. It ultimately became a public company with an Initial Public Offering (\textsc{IPO}) in 2020. 

Angel investors and VCs make many high-risk bets, and typically seek massive returns \cite{gompers2001venture}; as a consequence, they demand high-growth in very large markets. This tends to drive a ``growth at all costs'' mindset: for example, early Youtube grew exponentially by allowing copyright infringement \cite{gizmodo}. Often, multiple VCs will collectively imagine futures in which a certain technology becomes dominant \cite{beckert2016imagined}, producing a herding effect. Arguably, we see this currently among AI startups. For instance, a piece in \textit{The New York Times} reports on an AI startup that went from zero to a \$100M valuation in one week---due to multiple VCs rushing to the fund the new company \cite{nytai}.

As tech companies mature, however, additional expectations, and an expectation for steady growth, are placed on them \cite{kaplan2009should}. For example, as Reddit tried to raise a Series C funding round, prospective investors were concerned about the risk to the brand presented by Reddit's early tolerance of hate groups on the site (an artifact of earlier, high-growth VC interests) \cite{chandrasekharan2017you}. In the U.S., the \textsc{SEC} mandates that public tech firms like Alphabet, Meta, Apple, etc., release public, quarterly earnings reports \cite{sec}. This can drive more short-termism by large firms---as compared to early-stage startups---as the next quarterly report (and stock price movement it causes) is never far off.

\subsection{Capital and Computing Research}

An emerging line of critical scholarship engages with capital's impact on computing research. For example, in an essay, Whittaker explores how Big Tech companies manipulate academia to advance their interests and avoid regulation, often by funding research that frames issues as solvable technical problems \cite{whittaker2021steep}. Young, Katell, and Krafft argue that the enthusiastic participation of technology firms in the FAccT conference serves to certify their commitment to ethics while simultaneously silencing critical perspectives on the real-world consequences of AI harms \cite{young2022confronting}. They point to a study by Wilson et al. as an example of conflicted research: the authors, including employees of Pymetrics, received funding from the company whose algorithmic employment screening system they were analyzing \cite{wilson2021building}.

Phan et al. highlight the dilemmas faced by AI ethics researchers navigating the complex ``economy of virtue'' within Big Tech. They argue that researchers are often pushed into positions of ``working on the inside,'' ``working adjacently,'' or ``working outside,'' each with its own set of compromises and potential for co-option. Some concrete proposals have been made to mitigate these concerns: Young, Katell, and Krafft propose rigorous financial disclosure policies (the present work adopts this convention), greater participation from advocates and activists directly impacted by algorithmic systems, and promoting a culture of agonistic dissent \cite{young2022confronting}.

``Socialist pitches,'' \cite{lindtner2020prototype} such as those sometimes seen in open-source software, social design, and socially-inspired hackathons are not immune to these dynamics. Zhang and Carpano's analysis of the Chromium browser project demonstrates how seemingly open systems can be strategically leveraged to advance corporate interests and shape industry standards \cite{zhang2023chromium}. They argue that Google's control over Chromium, combined with its ``radical implementation-first'' approach, allows the company to exert significant influence over the direction of web development. Irani likewise argues that social design and socially-inspired hackathons can both be made to serve the interests of capital \cite{irani2019chasing,irani2019hackathons}. The broader political economy of technology is essential for comprehending the complex interplay of power, funding, and research within computing and computing research.

The present work builds on this line of scholarship by 1) introducing the concept of technological capture, and 2) applying it to CHI to analyze how it structures CHI research into different forms aligned with capital interests.
 
\section{Technological capture}

\begin{quote}
\emph{The value of a commodity is, in itself, of no interest to the capitalist. What alone interests him, is the surplus-value that dwells in it.}
--- Marx, \emph{Capital} \cite{marx2004capital}

\end{quote}

\vspace{2pt}

\noindent
A theoretical concept this paper contributes is the idea of \emph{technological capture}. Technological capture is when technological futures are captured by capital. Many possible technological futures are possible; however, in the process of technological capture, futures that do not align with capital interests are foreclosed upon.

The concept arises by way of analogy. In macro-economic contexts, industries employ lobbyists to bend regulatory agencies into alignment with their commercial interests. Banks lobby Congressional subcommittees to roll back protections put in place to limit the risk of systemic collapse \cite{hill}. Energy companies lobby the EPA to loosen restrictions on where oil can be extracted \cite{propublica}. Often, the firms are successful. The regulatory capacity of the agency becomes ``captured'' by the firms it was established to regulate. In the same way, technologicical futures are captured by capital in technological capture.

\subsection{Mechanisms}
Capital directs the shapes technologies are allowed to take. This happens via at least two different mechanisms. First, VCs play a significant role selecting for technologies that may present large rates of return. Second, in more established firms, public financial markets significantly shape technologies and their affordances. As considerably more has been written about public financial markets and firm decision-making (e.g., \cite{kaplan2009should}), we focus on VCs---VCs also more closely match the technology stage of much CHI research.

\vspace{8pt}
\noindent
\textit{\textbf{VCs.}} VCs play an outsize role in the tech industry relative to other industries; they also have outsize influence in which technologies make it out of labs. Most VCs are looking for so-called ``100x'' returns on technologies \cite{pg}. Paul Graham, the founder of YCombinator, writes: 

\begin{quote}
\textit{In purely financial terms, there is probably at most one company in each YC [YCombinator] batch that will have a significant effect on our returns, and the rest are just a cost of doing business ... I'm not saying that the big winners are all that matters, just that they're all that matters financially for investors.} \cite{pg}
\end{quote}

In this world, an investment is only worth making if it has the potential to yield a 100x return. It would make sense in the logic of rate of return to tolerate high failure rates to achieve it for the VCs' underlying investors---a common feature of venture capital. Therefore, a potential technological innovation must conform to the 100x rule; others are uninteresting to VCs. For example, new internet technologies that may dramatically improve the lives of hundreds of thousands of people are vastly inferior to attention-capturing technologies that can scale to global audiences.

This happens constantly in the tech industry. For example, capital has privileged surveillance capitalism over subscription-based internet platforms for decades \cite{zuboff}. Despite recent advances in federated learning \cite{li2020federated} that would unlock privacy-preserving AI systems, investors elevate ``massive model'' companies \cite{nytai}---perhaps anticipating a single winner in the future. These technological futures were possible, but they are unaligned with capital interests.

The process by which this happens is often clear and overt \cite{kupor2019secrets,thompson2022pilgrimage}. For instance, technologists take their early ideas for tech products to Sand Hill Road to pitch VC firms. There is a clear and explicit analysis of the benefits to capital interests in this exchange. Founders present pitch decks that focus on the capital interests of their new technology or technological platform: How big is the potential market? How big is the ``moat'' around the technology (i.e., how difficult would it be for a competitor to displace you)? At what valuation is the nascent firm seeking investment, and what does that valuation mean for an investor's rate of return?

\section{Technological capture and CHI}

How does technological capture affect CHI? We argue that CHI has inherited capital logics and the resulting technological capture via many vectors:

\begin{enumerate}
    \vspace{6pt}
    \item[] \textbf{Many founders of the field held positions in the early tech industry.} CHI had considerable early influence from the tech industry. Founders of the field held key positions at Microsoft, Apple, PARC, DEC, and Sun, among others \cite{grudin2008moving}. In a history of HCI, Jonathan Grudin writes:

    \vspace{6pt}
    \begin{quote}
    \textit{[A history of HCI] focuses on technologies and practices as they became widely used, as reflected in the spread of [industry's] systems and applications. This was often paralleled by the formation of new research fields, changes in existing disciplines, and the creation and evolution of professional associations and publications.} \cite{grudin2008moving}
    \end{quote}
    
    \vspace{6pt}
    \item[] \textbf{Revolving doors between faculty and industry.} It is not uncommon for CHI faculty to move between the academy and industry positions---and sometimes hold both simultaneously \cite{cra}. Written in part by CHI community members, a recent Computing Research Association\footnote{CRA is a non-profit that represents 250 computing research organizations in North America, including research-active U.S. Computer Science departments.} (CRA) report states: 

    \vspace{6pt}
    \begin{quote}
        \textit{We observe significant increases in the level of interaction between professors and companies, which take the form of extended joint appointments ... This increasing connection ... has the potential to change (either positively or negatively) numerous things, including: 1) The academic culture in computing research universities; 2) The research topics that faculty and students pursue} \cite{cra}.
    \end{quote}

    \vspace{6pt}
    \item[] \textbf{Research projects become startups.} Technologies that start in academic labs become startups. Universities have offices dedicated to ``tech transfer,'' the process transitioning a research-level prototype or technique into a product that can be sold. For example, The University of Michigan's office of Innovation Partnerships \cite{umich} offers the following service to faculty:

    \vspace{6pt}
    \begin{quote}
        \emph{Our experienced team of licensing professionals is available to support you on your innovation journey. Connect with us to discuss your research, submit an invention disclosure, and begin to map a commercialization and impact strategy.}
    \end{quote}

    \vspace{6pt}
    Most R1 universities in the United States have similar divisions. A number of successful startups trace their origin to research published in CHI: Net Perceptions was an early startup that transferred recommender systems research to Amazon \cite{netperceptions}; many faculty at universities with significant CHI representation, including the University of Washington and Carnegie Mellon, have startups that commercialize CHI research (e.g., \cite{niki,lva,patel}).

    \vspace{6pt}
    \item[] \textbf{Industry steers funding.} Companies often offer funding for research that aligns with their vision of computing. For example, Meta will frequently run funding competitions for ``research aligned with our mission'' \cite{metaawards}. It is arguable whether the purpose of such calls is to direct actual research, to cultivate relationships with certain faculty and their students, or to advance favorable public relations campaigns---perhaps all three. This process can also take subtle forms. For example, when the \textsc{NSF} announced its Program on Fairness in Artificial Intelligence, Amazon was a founding partner \cite{nsfai}.

    \vspace{6pt}
    \item[] \textbf{Corporate sponsorship of the conference.} Tech companies routinely sponsor the conference, and have consistently over its lifetime. At the most recent \textsc{CHI} conference at the time of this writing, \textsc{CHI} 2024, corporate sponsors included Apple, Google, Meta, Microsoft, and Adobe \cite{sponssors}. 

    \vspace{6pt}
    \item[] \textbf{Implications for design.} One of the basic values of CHI \cite{dourish2006implications}, often shorthanded as the ``Implications for Design'' section, is a mandate to make research informative for tech products operated by tech firms. I.e., why does it matter for product? Implications for design can be cast as a direct through line between research design and why it matters for return on capital.

    \vspace{6pt}
    \item[] \textbf{CHI works on systems that have already been captured.} Finally, we inherit technological capture by studying and building on systems that have already been captured by capital. At CHI, we often work on systems that already exist---many of which are owned and operated by tech firms. We ``plug into'' the capital logics that give way to technological capture. For example, if we want to improve exposure to misinformation on Facebook, we have to ``sell'' the research within the logic of surplus value: any intervention must be fast, cheap, and not interfere with the main profit-generating parts of Facebook (e.g., \cite{jahanbakhsh2021exploring,resnick2023searching}).
    
\end{enumerate}

\section{How Technological Capture Structures CHI Research} \label{sec:forms}


\noindent
Next, we attempt to derive a theoretical framework for how technological capture structures CHI research. It would be ill-advised to develop one that encompasses all research \textsc{CHI} has ever published; the objective is to develop a framework that can contain broad arcs and archetypes of \textsc{CHI} research. We introduce a framework with four parts:  technological capture structures \textsc{CHI} as \textit{market-creating}, \textit{market-expanding}, \textit{market-aligned}, and \textit{externality-reducing} forms.

\subsection{Method}

While large and complex, \textsc{CHI} presents itself to aspiring authors as comprising a set of ``subcommittees'' \cite{chi24}. A subcommittee, in turn, states its values and selects previously-accepted papers as exemplars. Where possible, we will draw upon this decomposition of the field, and the ways subcommittees state their values, in our analysis. We employ frame analysis as the method for analyzing subcommittees statements of values and their representative papers \cite{greene2019better,goffman1974frame}. We use the current CHI 2025 subcommittee pages, and the papers they reference, as our data \cite{chi24}.

There is a dynamism to CHI that this approach does not fully capture. For example, subcommittees change over time both in their focus and in their composition. New ones are added, and older ones are sometimes removed or collapsed with others. Our approach is a single snapshot of the conference, used as reflection of the ties between types of CHI research and capital.

\subsection{First form: Market-creating}
Capital requires new firms and products be created. Many of the biggest companies by market capitalization were founded within living memory (e.g., Apple, Meta, Alphabet). The process of creative destruction \cite{elliott1980marx} that finds and elevates companies like these requires a constant churn of new firms. Often, the most disproportionate returns come from these firms.

At the same time, new raw technologies become available. The number of pixels that manufacturers can place in one inch may dramatically increase. Drones suddenly become light enough and cheap enough to carry objects around households. Video capture technologies plummet in their cost to manufacture, leading to their placement all over living spaces. On their own, however, these technological changes do very little \textit{for people.} Historically, it has fallen to communities like \textsc{CHI} to imagine, develop, and test ways to use these technologies toward human ends---even if the \textsc{CHI} community did not participate in the original creation of the new tech. For example, both the UIST community and the NSF claim credit for foundational research that brought iOS devices to market \cite{NSF}. These innovations created new markets for Apple---the gains from which are now largely concentrated in the hands of Apple's shareholders. 

The Blending Interaction subcommittee solicits ``novel interactive systems and enabling contributions'' \cite{chi24}. The Developing Novel Devices subcommittee ``focuses on advancing interaction through developing novel hardware and physical devices'' that demonstrate ``showcased relevance through example applications'' \cite{chi24}. The Interacting with Devices subcommittee solicits ``enabling interactions using different modalities, such as touch, gestures, speech \& sound.'' Specifically, that subcommittee states: 

\begin{quote}
\textit{Contributions will be judged based how well a proposed approach solves a significant existing problem or how well it \textbf{opens new and compelling opportunities} [emphasis added] for interactions. The novelty of the interaction, its design rationale, and evaluations demonstrating improvements over existing interaction techniques are particularly well suited for this committee.} \cite{chi24}
\end{quote}

\noindent
These values closely mirror capital's interest in creating new products and firms. It is from ``enabling contributions'' that startups can be created to manufacture and sell a technology.

\vspace{6pt}
\noindent
\textbf{\textit{Representative subcommittees.}}
Interacting with Devices: Interaction Techniques \& Modalities; Developing Novel Devices: Hardware, Materials, and Fabrication; Blending Interaction: Engineering Interactive Systems \& Tools

\subsection{Second form: Market-expanding}
Capital requires that customer bases expand; they must not contract. Once a firm finds traction with a technology, and has customers, capital demands that those customers grow. A firm that has a strong base in the United States must look to expand overseas. A firm, such as Meta or Alphabet, reliant on digital advertising, must increase advertisers buying ads to put in front of users. Moreover, when customer bases contract, firms are often punished, either in public financial markets or by VCs.

It is from this capital value that we receive a great deal of UX and ``understanding users'' research, from which we make implications for design. In its description, The User Experience subcommittee solicits research contributions that ``make technology more \textit{useful, usable}, and \textit{desirable} [emphasis added].'' The Understanding People subcommittee seeks ``improved understanding of people and/or interactional contexts.'' These findings can improve people's experiences with tech, yielding longer sessions, more page views, and fewer customer de-activations.

Research from these subcommunities has studied what features, activities, and interactional contexts lead to higher ``engagement'' (e.g., \cite{bakhshi2014faces,grinberg2016changes}). Work has extensively studied the cultural contexts surrounding use of digital technologies in countries outside the United States (e.g., \cite{cheng2021country,nouwens2017whatsapp}). These contexts often surprise the largely Western designers that originally envisioned the tech, leading to issues around adoption and continued use outside the U.S.; however, many of these same markets represent growth opportunities for Western tech firms, and demand adaptation.

These values are direct complements of capital interests to please, retain, and expand customer bases. It also echoes work that argues that CHI's elevation of \textit{usefulness} ``masks various forms of violence and injustice'' \cite{lin2021techniques}.

\vspace{6pt}
\noindent
\textbf{\textit{Representative subcommittees.}} Interaction Beyond the Individual; Understanding People; User Experience and Usability

\subsection{Third form: Market-aligned}
Capital has outsize interest in particularly lucrative markets. Capital will follow and respond to the opportunity to extract surplus value from these certain highly-lucrative, or potentially highly lucrative, markets. If a market or market segment exhibits these properties, we should expect over-representation there from capital. 

Examples include health and aging populations. The global population is rapidly aging---especially in wealthy countries such as the USA, Japan, and across the EU. The United Nations reports that ``virtually every country in the world is experiencing growth in the number and proportion of older persons in their population'' \cite{un}: it estimates that ``1 in 6 people in the world will be over the age 65 by 2050, up from 1 in 11 in 2019'' \cite{un}. The health market, likewise---both worldwide and especially in the United States---is highly profitable: e.g., price inflation far exceeds inflation in other areas of the economy. For example, according to the Peterson Center on Healthcare and KFF: 

\begin{quote}
    \emph{Since 2000, the price of medical care, including services provided as well as insurance, drugs, and medical equipment, has increased by 114.3\%. In contrast, prices for all consumer goods and services rose by 80.8\% in the same period.} \cite{kff}
\end{quote}

\noindent 
Both of these markets---one in terms of raw customer numbers, and the other in terms of price growth---represent large opportunities for capital investment. Likewise, we see subcommittees dedicated to them at CHI. The Accessibility and Aging subcommittee calls for research dedicated to ``technology designed for or used by people in the later stages of life'' \cite{chi24}. The Health subcommittee calls for research contributing a ``clear and novel contribution to HCI in terms of our understanding of people’s interaction with technology in a healthcare context'' \cite{chi24}.

These objectives parallel capital values to maximize rates of return by targeting large and lucrative markets.

\vspace{6pt}
\noindent
\textbf{\textit{Representative subcommittees.}} Accessibility and Aging; Games and Play; Health

\subsection{Fourth form: Externality-reducing}
Finally, in its exhaustive search for surplus value, capital creates negative externalities. A negative externality is a cost imposed by one actor on other actors, but not paid for by the originator.

For example, many internet platforms hesitate to put any barriers in place between their users and posting: posts create content that others can view, without which ads cannot be shown. From this, internet communities downstream receive misinformation, hate speech, and harassment. The research community has expended considerable effort to algorithmically identify, triage, remove, and safeguard people from these types of content (e.g., \cite{epstein2020will,juneja2021auditing,ma2022brush,mahar2018squadbox,thomas2022s,varanasi2022accost}). Typically, a value in these types of studies is cost-effectiveness---either in their systems or their implications---as it is presumed that internet platforms will not put considerable resources into the problems. Similarly, whole conferences like FAccT have emerged to mitigate problems of bias and fairness downstream of fast-moving AI and algorithmic systems in industry. An example is that it now falls to the academic community to study and rectify the false information from ChatGPT (e.g., \cite{donia2022normative,zhou2023synthetic}). Another form is when---in the pursuit of dominant, wealthy customers---other users are marginalized by tech. It falls to CHI to build new ways to make existing systems accessible, or to adapt identity systems for marginalized users, such as transgender people (e.g., \cite{haimson2015disclosure,scheuerman2018safe}). Externality-reducing research is different from the previous three forms: it is often not directly on the critical path of maximizing rates of return. This is also where we sometimes see research that makes \textit{socialist pitches}, like ``tech that will make the world a better place'' \cite{lindtner2020prototype}.

\vspace{6pt}
\noindent
\textbf{\textit{Representative subcommittees.}} Privacy and Security; Specific Applications Areas; Accessibility and Aging; Interaction Beyond the Individual; Critical Computing, Sustainability, and Social Justice

\subsection{CHI research outside capital logics}
As Figure \ref{fig:capture-chi} suggests, many CHI research projects do not share values with capital interests via technological capture. They would not fall into one of the four forms above. CHI has published research that employs feminist philosophies to interrogate civic participation \cite{meng2019collaborative}. CHI has published work that examines the ways tech helps or hinders community activists \cite{asad2015illegitimate}. The community has also explored anti-capitalist design frameworks, such as social justice-oriented interaction design \cite{dombrowski2016social}. A capital-oriented theory of CHI also leaves out research on Wikipedia (e.g., \cite{smith2020keeping,vincent2018examining}) and open source communities (e.g., \cite{qiu2023climate}).

Yet, also as Figure \ref{fig:capture-chi} intimates, we theorize that the fraction of work this represents is proportionally small. For example, a search on Google Scholar for CHI research that mentions ``Facebook'' returns 2,760 results \cite{gs1}; the same search for ``Wikipedia'' returns 795 results \cite{gs2}---despite the relative openness of Wikipedia to academic research.
 
\section{Arguments against technological capture at CHI} \label{sec:counter}
There are arguments against the technological capture theory presented in this paper. They should be examined closely. First, one could argue that CHI developed independently enough from capital concerns, and that it has values outside capital ones. Second, one could argue that the field has grown beyond this capital-derived set of values---that much recent work is focused on critical perspectives, social justice, and anti-racist value systems.

First, it could be that CHI developed independently enough from industry and its underlying capital interests to develop its own value systems. There are a number of values CHI holds which do not come from capital concerns. One example is \textit{rigor}. For example, the Understanding People subcommittee states that papers will be selected for ``their rigor, significance, originality, validity'' \cite{chi24}. A good deal of early CHI research grew out of psychological traditions that prized experimental and statistical rigor. CHI adopted these values as well \cite{grudin}; and, certainly few would ascribe \textit{rigor} to the interests of capital underlying the field. In this view, CHI may share some of its values with capital, but it has its own distinct ones that extend or supersede capital interests.

We do not disagree with this counter-argument in its entirety. We would not argue that all of CHI's values are inherited from capital interests; instead, we would argue that many of its most essential ones are.

The second counter-argument to this framing concerns the growth of the field. In this argument, perhaps it is acknowledged that the field grew alongside industry and even inherited capital interests from it, but that the field has transcended that early history. For example, CHI now has a subcommittee named \textit{Critical Computing, Sustainability, and Social Justice}, which ``welcomes HCI research connected to themes of social justice, global sustainability, critical-reflective research practice, artful and aesthetic experiences, and critical computing'' \cite{chi24}. Few would connect these values with the values of capital. 

Yet, the Critical Computing, Sustainability, and Social Justice subcommittee was only introduced in 2021 \cite{chi21}. However, it becoming distinct (i.e., most of its core constituents were spread throughout the rest of the conference before it existed) does allow us to examine its relationship to the broader whole: it is fair to conclude that it is quite small. Previous to its introduction in 2021, one might presume that similar papers were directed elsewhere, perhaps to the \textit{Design} subcommittee. However, bibliometric analyses of the CHI conference over time show that these papers remain a small fraction of the overall total---even when considering main themes \cite{pohl2019we}.

To summarize, we would argue that the many, but not all, of CHI's most essential values have been inherited from capital interests, and that countervailing values remain proportionally small and nondominant. 

\section{Conclusion}
Increasingly within the critical constituencies of \textsc{CHI} the question is raised: \textit{Why are we doing this work?} Why develop new ways to ship work down a wire, allowing work to be crowdsourced all over the globe \cite{kittur2013future}? Why create new ways to interact in virtual reality (e.g, \cite{benko2016normaltouch,cheng2015turkdeck})? Why build robots to care for aging populations (e.g., \cite{carros2020exploring,marchetti2022pet})? Why study how to make self-driving cars easier to drive (e.g., \cite{currano2021little,du2020evaluating})? Why create hate speech detection models for internet platforms (e.g., \cite{chandrasekharan2019crossmod})?

Our answer is this: because the research is structured by capital. It creates new products. It creates new firms. It retains or expands customers bases. It places technology within particularly lucrative markets. It helps mitigate externalities with low-cost techniques---perhaps allowing firms to sidestep regulation.


\subsection{Implications}
Our primary purpose in writing this piece is largely explanatory, rather than normative. Our argument aims to show that many of CHI's essential, underlying values mirror corresponding capital interests via technological capture. We are not, however, arguing that any industrial connection or relevance is unequivocally bad. One could argue that some level of connection and relevance to industry is healthy. Without it, many other academic fields have languished by comparison with CHI. For instance, The Great Hall at the National Academy of Sciences in Washington, D.C., has an inscription that reads: 

\begin{quote}
    \textit{To science, \textbf{pilot of industry} [emphasis added], conqueror of disease, multiplier of the harvest, explorer of the universe, revealer of nature's laws, eternal guide to truth.} \cite{nas}
\end{quote}

Arguably, science has played an important role relative to industry in modern societies. Yet, as an academic discipline, we should not passively receive capital interests as our academic or scientific values.

As a field, we should have dialogues about where our values come from, and what we ultimately want them to be. If we want to decouple from capital, we should consider academic and organizational changes that would incent---to borrow an economic mechanism---work that grows CHI further outside the shadow of technological capture. For example, we want may consider how we can foster more self-sustaining systems like Turkopticon \cite{irani2013turkopticon} that reject capital interests. We may consider adopting a new norm: an \emph{Implications for Society} instead of \emph{Implications for Design}. We may consider special calls, with special incentives, for work on open culture and open source. CHI is also particularly well-placed to imagine new ways to collectively assemble, manage, and own capital---in opposition to now-dominant cryptocurrency approaches.

Perhaps paradoxically, it may be here---beyond the the shadow of technological capture---that the greatest technological innovation is possible. One could argue that tech and research within the bounds of technological capture will necessarily happen. Systems and research outside the boundaries of technological capture will instead need substantial support---financial, infrastructural, etc.---to launch. Afterward, how do we keep them going? CHI and its sister communities are particularly well-placed to consider such questions.

\section{Limitations}
This work has a number of limitations. Perhaps first and foremost, the work is situated in a Global North, and particularly North American, context. This work also does not account for the dynamism of CHI, such as changes to subcommittees over time. Because there is comparatively more written about public companies and public financial markets, the present work primarily focuses on VCs and does not discuss in detail the mechanisms by which public financial markets shape technology. Finally, this work is largely theoretical and explanatory, rather than normative: we do not make many prescriptions for change. Future work could expand upon and improve these areas.

\section{Funding disclosure} \label{sec:funding}
We are employed at a large North American university. In addition to our academic salaries, we have received research funding from government agencies (NSF, DARPA), private foundations (SSRC, Craig Newmark Philanthropies), and from large tech firms (Alphabet, Yahoo!, Microsoft and Meta) over our careers. We have never accepted grants from companies or agencies that come with publication oversight.

The following conforms with FAccT’s Conflict of Interest policy \cite{coi}. Our employer and our funders played no role in designing or conducting this work. They played no role in accessing or collecting data. They played no role in data analysis or interpretation. They played no role in preparing, reviewing, or approving the manuscript. They played no role in the decision to submit the manuscript for publication. We have no financial ties to any entity affected by publication of the paper.

\section{Acknowledgements}
We thank various colleagues who provided feedback on early drafts of this work.

\bibliographystyle{ACM-Reference-Format}
\bibliography{references}

\end{document}